\documentclass[reprint,
 superscriptaddress,
 secnumarabic,
 amsmath,amssymb,
 aps,
 prl,
 nobibnotes
]{revtex4-2}

\usepackage{graphicx}
\usepackage[colorlinks=true,linkcolor=blue,citecolor=blue,urlcolor=blue]{hyperref}
\usepackage{svg}
\usepackage{dcolumn}
\usepackage{bm}
\usepackage{braket}
\usepackage{cancel}
\usepackage[normalem]{ulem}

\usepackage[english]{babel}

\begin{document}

\preprint{APS/123-QED}

\title{Quantum Dispersive Waves and Multimode Squeezing in \\Pure-Kerr Parametrically Driven Cavity Solitons}

\author{Rafael Romero Mendez}
\thanks{These authors contributed equally}
\affiliation{Institute for Physical Science and Technology, University of Maryland, College Park, Maryland 20742, USA}
\affiliation{Department of Physics, University of Maryland, College Park, Maryland 20742, USA}
\author{Sashank Kaushik Sridhar}
\thanks{These authors contributed equally}
\affiliation{Department of Mechanical Engineering, University of Maryland, College Park, Maryland 20742, USA}
\author{Samyak Gothi}
\affiliation{Institute for Physical Science and Technology, University of Maryland, College Park, Maryland 20742, USA}
\author{Pradyoth Shandilya}
\affiliation{University of Maryland Baltimore County, Baltimore, Maryland 21250, USA}
\author{Yichen Shen}
\affiliation{Department of Mechanical Engineering, University of Maryland, College Park, Maryland 20742, USA}
\author{Curtis Menyuk}
\affiliation{University of Maryland Baltimore County, Baltimore, Maryland 21250, USA}
\author{Avik Dutt}
\email{avikdutt@umd.edu}
\affiliation{Institute for Physical Science and Technology, University of Maryland, College Park, Maryland 20742, USA}
\affiliation{Department of Mechanical Engineering, University of Maryland, College Park, Maryland 20742, USA}
\affiliation{National Quantum Laboratory (QLab) at Maryland, College Park, Maryland 20742, USA}

\date{\today}

\begin{abstract}

Parametrically driven cavity solitons (PDCS), unlike single-pumped cavity solitons, are localized optical pulses arising from parametric processes. These cavity solitons, recently discovered in pure-Kerr media, offer great promise for nonlinear dynamics studies and metrology. Here, we present the first multimode quantum description of pure-Kerr PDCS. In the below threshold regime, we verify single- and two-mode squeezing, while above threshold we uncover novel ``quantum" dispersive waves—the quantum analog of soliton Cherenkov radiation. Besides revealing these unexplored quantum properties, we show that PDCS generates up to 20 dB of squeezing, only limited by overcoupling and intrinsic losses for experimentally routine parameters. We therefore provide a pathway to observe strong multimode quantum noise reduction in these systems.

\end{abstract}

\maketitle



\textit{Introduction}—Parametrically driven cavity solitons (PDCS) are a special kind of optical dissipative and temporal cavity solitons (CSs) arising in driven-dissipative resonators \cite{mecozzi_long_1994, longhi_ultrashort_1995, de_valcarcel_phase_2013, englebert_parametrically_2021}.
A pure-Kerr PDCS \cite{moille_parametrically_2024} can be excited by bichromatically driving two continuous-wave (CW) lasers, leading to the formation of a sech-shaped frequency comb. Unlike single-pumped dissipative Kerr solitons (DKSs), the PDCS spectrum lacks a pump at the center, allowing for easier spectral extraction of the comb. PDCSs thus promise a wealth of interesting applications in metrology, and fundamentally unique perspectives on the nonlinear dynamics of solitons beyond DKS \cite{moille_self-aligned_2026}. 
\begin{figure*}
\includegraphics[width=\textwidth]{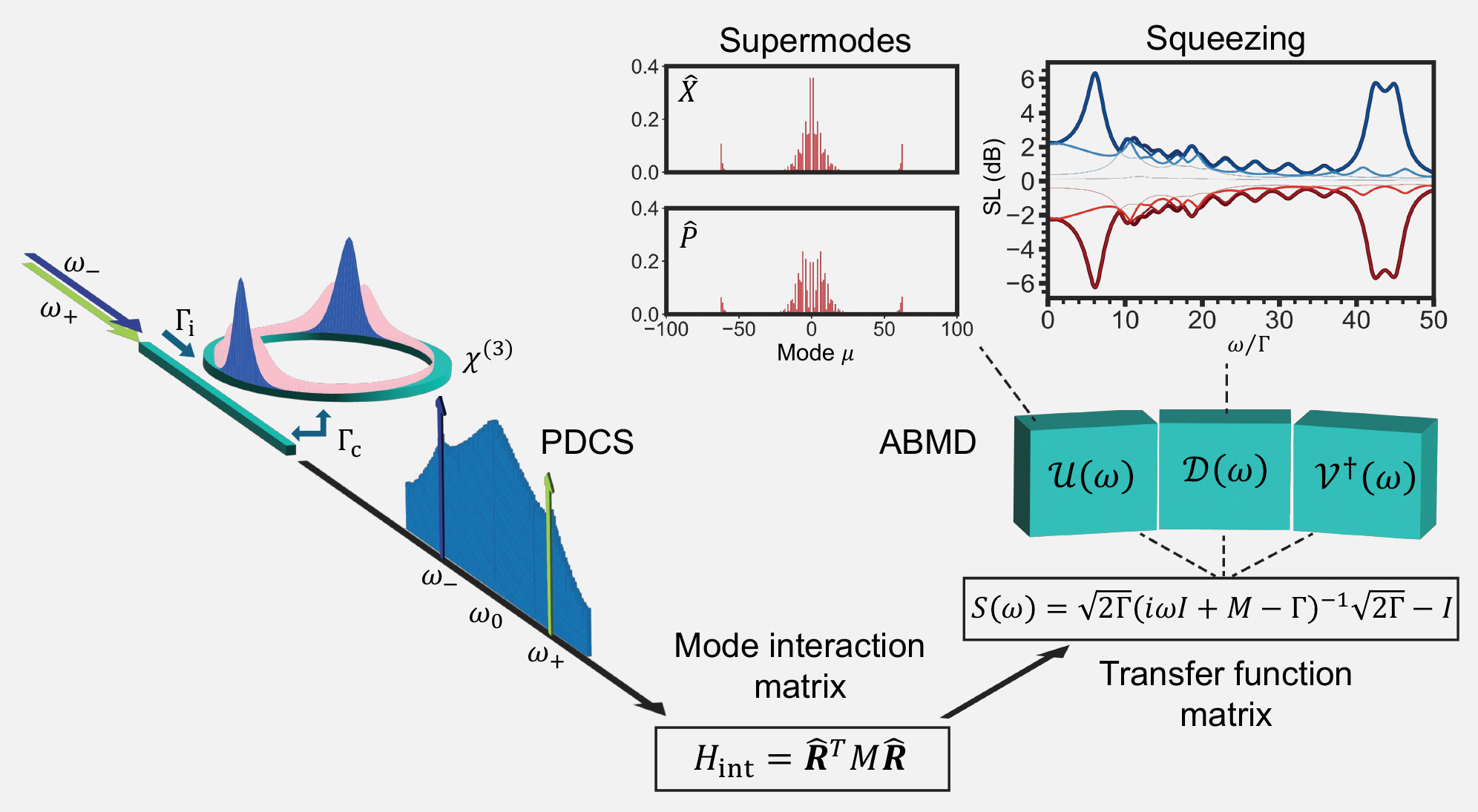}
\caption{\label{fig:wide}Schematic depiction of the quantum multimode analysis of a pure-Kerr parametrically driven cavity soliton (PDCS). The microresonator with intrinsic loss $\Gamma_i$ is coupled to a waveguide with rate $\Gamma_c$. It is pumped bichromatically (\(\omega_{\pm}\)) with opposite phases, inducing two coherent soliton pulses in temporal domain (intensity in blue) with quantum noise temporal envelopes (pink). The resulting PDCS spectrum is used to obtain the cavity interaction Hamiltonian in terms of boson quadratures \(\hat{\mathbf{R}}=(\hat{\mathbf{X}}|\hat{\mathbf{P}})^T\) and the mode interaction matrix \(M\). Using input-output relations and the Heisenberg-Langevin equation in Fourier domain, the input and output quadratures can be related via \(S(\omega)\), which can be decomposed using the analytical Bloch-Messiah decomposition (ABMD) \cite{gouzien_morphing_2020}, hence obtaining the multimode squeezing levels (SL) with their associated supermodes.}
\label{fig:schematic}
\end{figure*}

In concurrence, studies of the quantum properties of Kerr microcombs have led to the generation of quantum squeezed states in Kerr microresonators \cite{yang_squeezed_2021, jahanbozorgi_generation_2023, jia_continuous-variable_2025, wang_large-scale_2025, shen_highly_2025}. Early theoretical studies involving noise in standard Kerr microcombs have used pairwise mode analysis \cite{herr_universal_2012, chembo_quantum_2016}, focusing on the signal-idler basis, and multimode squeezing analysis of bulk parametric oscillators \cite{de_valcarcel_multimode_2006}. Recently, multimode quantum correlations have been analyzed in femtosecond pulses in nonlinear fibers \cite{uddin_noise-immune_2025} and soliton crystal states in Kerr microcombs \cite{guidry_multimode_2023}.
Such squeezed frequency combs can host highly entangled continuous-variable states \cite{moran_experimental_2014,pfister_continuous-variable_2019,yang_squeezed_2021, jahanbozorgi_generation_2023, wang_large-scale_2025, jia_continuous-variable_2025,shen_highly_2025,li_perfect_2025, Lustig2025}.  

The question arises thus: beyond revealing quantum phenomena similar to standard Kerr microcombs, can a quantum analysis of PDCS predict fundamentally novel states and behavior? 
From an application perspective, this analysis merits study because PDCS formation is fundamentally underpinned by degenerate phase-sensitive amplification, with the potential to practically isolate a single-mode squeezed vacuum from the far-separated pumps --- a regime impossible to achieve in single-pumped microcombs, and difficult to filter in standard dual-pumped Kerr microcombs \cite{Zhang_squeezed_2021, Zhao_near-degenerate_2020}. The importance of single-mode squeezing is underscored by its extensive use in protocols for quantum sensing and quantum information processing \cite{aasi_enhanced_2013,nehra_few-cycle_2022,Frascella_Multimode_2021,shaked_lifting_2018,virally_enhanced_2021,moille_parametrically_2024}.

Here, we reveal the multimode squeezing behavior of pure-Kerr PDCSs. The morphing supermode analysis \cite{gouzien_morphing_2020} of the PDCS reveals highly squeezed states saturating the limits on quantum noise reduction set by the overcoupling and intrinsic loss across strikingly different regimes of the system's classical phase diagram. Below the soliton threshold, we see clean single- and two-mode squeezing associated with degenerate and non-degenerate optical parametric amplifiers (OPA) respectively \cite{wolinsky_quantum_1988,drummond_correlations_1990,chembo_quantum_2016,Zhao_near-degenerate_2020, Zhang_squeezed_2021}. Above threshold, we identify the remarkable emergence of ``quantum" dispersive waves (QDWs) for the first time, in analogy with classical dispersive waves in nonlinear optical systems \cite{PhysRevA.41.426, PhysRevA.51.2602, brasch_photonic_2016}. We identify regimes where the dominant squeezed supermode is spectrally localized at the QDWs, which we term ``pseudo" two-mode squeezing, and also confirm with the intra-cavity temporal noise profile. Note that a recent complementary study has predicted favorable performance of pure-Kerr PDCS compared to DKS in terms of noise characteristics, fundamental quantum-limited timing jitter and two-mode squeezing ~\cite{Shamailov2026PDCSNoise}. 
Taken together, these analyses shed light on fundamentally new quantum states of light in parametrically driven Kerr solitons, with potential applications in quantum sensing and frequency-domain information processing \cite{kues_quantum_2019, pfister_continuous-variable_2019, Zhang_on-chip_2023, wen_polarization-entangled_2023,herman_squeezed_2025, renault_end-to-end_2025}. 

\textit{Classical phase diagram}— We firstly review the mean-field dynamics of bichromatically driven Kerr resonators, which are numerically modeled by the extended Lugiato-Lefever equation \cite{Taheri2017}. This model permits the simulation of PDCS in resonators pumped by two coherent CW fields \(E_\pm\) at pump frequencies \(\omega_\pm\), forming a coherent frequency comb \(E_0\) at the signal frequency \(\omega_0 \approx (\omega_+ + \omega_-)/2\) via non-degenerate four-wave mixing (FWM) interactions (see Fig.~\ref{fig:schematic}). Under specific conditions on \(E_\pm\), this can be approximated by the (normalized) parametrically driven nonlinear Schr\"odinger equation (PDNLSE) \cite{moille_parametrically_2024}: 
\begin{equation}
    \frac{\partial E_0}{\partial t} = 
    \left[-1 + i\left(|E_0|^2 - \Delta_{\text{eff}}\right)  
    -i\hat{d}_\text{int}\left( \mu \right)\right] E_0 
    + \nu E_0^* 
\label{eq:pdnlse}
\end{equation}
Here, \(\hat{d}_{\text{int}}(\mu)=\sum_{n\ge 2}\frac{d_n}{n!}(\mu)^n\) is the normalized integrated dispersion of the cavity resonance, with $d_n = t_R D_n/ \Gamma$ where $D_n$ is the Taylor expanded component of integrated dispersion ($D_\mathrm{int}$) \cite{brasch_photonic_2016}. The grid is labeled by \(\mu\) as the mode number relative to \(\omega_0\), $t_R$ is the round trip time and $\Gamma$ represents the total amplitude decay per round trip. Importantly, \(\Delta_{\text{eff}}\) is the effective detuning, dependent on the detunings of \(E_{\pm}\) and the intensity \(|E_{\pm}|^2\) due to cross-phase modulation, and \(\nu = 2iE_+ E_- \) is the parametric drive coefficient. To ensure phase matching of the degenerate FWM process and consequent PDCS formation, the integrated dispersion must be anomalous around the central frequency \(\omega_0\) (\(d_\mathrm{int}>0\)) and near-zero at \(\omega_{\pm}\) \cite{moille_parametrically_2024}. The latter condition implies that at least one higher-order dispersion coefficient must be negative (in our case, \(D_4<0\)).

\begin{figure*}
\includegraphics[width=\textwidth]{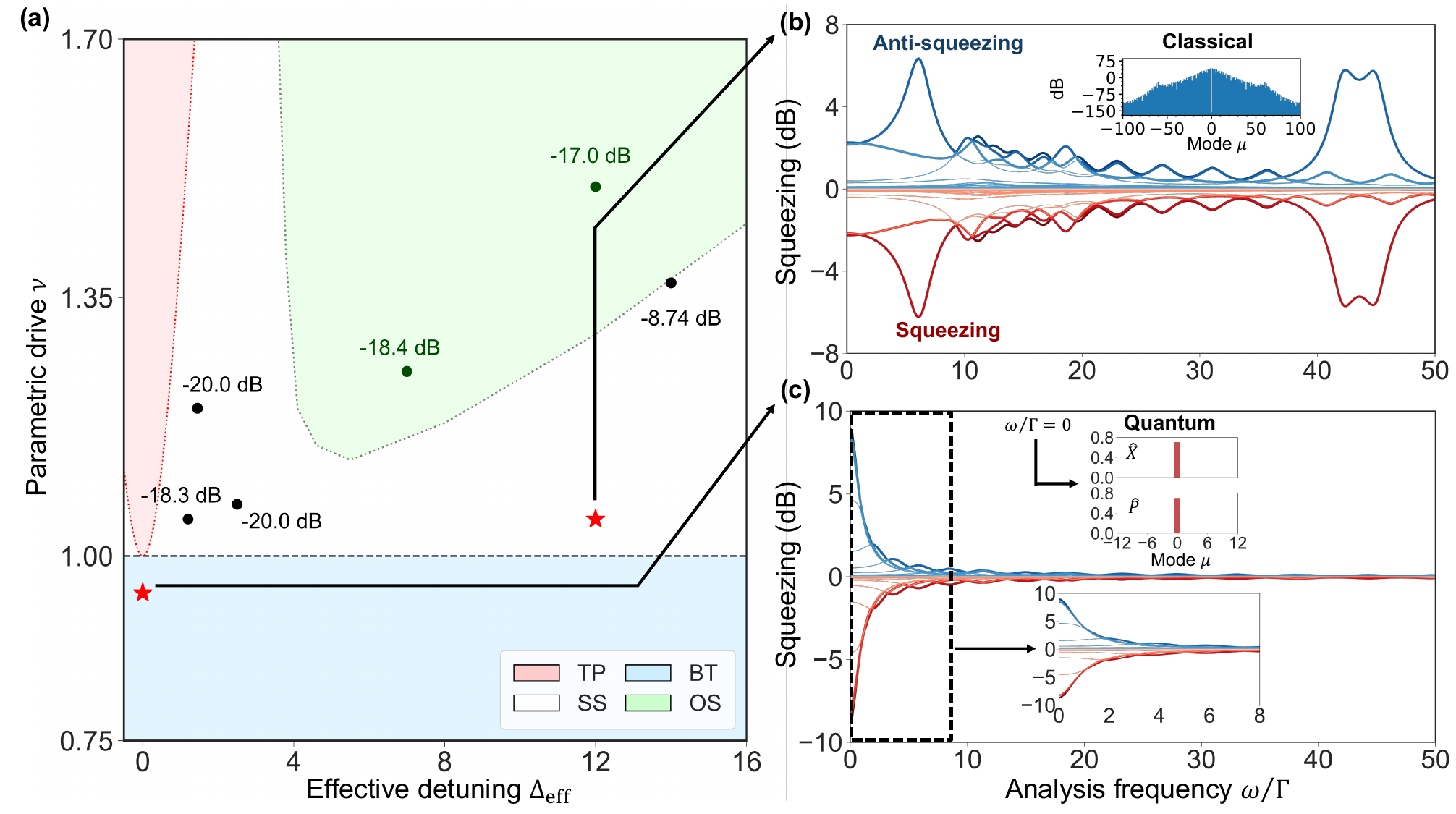}
\caption{\label{fig:wide} Classical phase diagram and quantum multimode analysis of the parametrically driven nonlinear Schr\"odinger equation (PDNLSE). (a) PDNLSE phase diagram in the \((\Delta_{\text{eff}},\nu)\)-space with representative values of highest squeezing across different dynamical regimes. Above threshold Turing Pattern (TP) states are in red; the green (Hopf) bifurcation curve corresponds to Oscillatory Soliton (OS) states; and Stable Soliton (SS) states are in white. The Below OPO-Threshold (BT, in blue) region shows no solitons. (b) Squeezing spectrum for a SS state \((\Delta_{\text{eff}}=12, \nu = 1.05)\). Inset: classical PDCS spectrum at the corresponding parameters. (c) Squeezing spectrum for a BT state \((\Delta_{\text{eff}}=0.0, \nu = 0.95)\). Top inset: degenerate single-mode supermode corresponding to the highest squeezing at \(\omega/\Gamma=0\). Bottom inset: Zoomed in squeezing spectrum showing the competing phase-matching between multiple pair-wise supermodes.}
\label{fig:phase_diagram}
\end{figure*}

Note that Eq.~\eqref{eq:pdnlse} yields a phase diagram in the \((\Delta_{\text{eff}},\nu)\)-parameter space. This phase diagram has been used to explore the stability of different soliton regimes for fiber-ring cavity optical parametric oscillators (OPOs) \cite{BONDILA1995314, englebert_parametrically_2021}. Using parameters from a recent experiment \cite{moille_parametrically_2024} (see S.1.), we simulate a PDCS formed by initializing an antisymmetric soliton pair in the \((\Delta_{\text{eff}}, \nu)\)-parameter space. Fig.~\ref{fig:phase_diagram}(a) shows the phase diagram of Eq.~(\ref{eq:pdnlse}) for selected parameter values, with approximate bifurcation boundaries based on \textcite{englebert_parametrically_2021}. We can divide this into two main regions, above-threshold (\(\nu>1\)) and below-threshold (BT) states (\(\nu<1\)). Above threshold, we focus on two stable regions: Oscillatory Soliton (OS) and Stable Soliton (SS), and proceed to analyze their multimode squeezing patterns in contrast with the below-threshold states.

\textit{PDCS quantum multimode analysis}—The evolution of the intra-resonator quantum field can be described in terms of bosonic operators \(\hat{a}_n\) associated with the resonant modes, following the standard boson commutation rules $\left[\hat{a}_j, \hat{a}_k^\dagger\right] = \delta_{jk}$ and $\left[\hat{a}_j, \hat{a}_k\right] = 0$.  Along with the two main pumps assumed to be placed at $\mu=\pm63$, the mean-field steady-state amplitudes ($A_m, A_n$ below) obtained from Eq.~(\ref{eq:pdnlse}) also act as pumps that stimulate FWM interactions, while their quantum behavior is ascribed to the bosonic operators, giving us a linearized Hamiltonian that dictates the dynamics of intracavity modes \cite{guidry_multimode_2023, gouzien_hidden_2023}: 
\begin{eqnarray}
    \nonumber H_\text{int} &=& \frac{\hbar g_0}{2}\sum_{m,n,j,k}\Big[4\delta_{(j-k+m-n)}A_m^*A_n\hat{a}_j^\dagger\hat{a}_k \\ 
    \nonumber &\ & +\,\delta_{(j+k-m-n)}\{A_mA_n\hat{a}_j^\dagger\hat{a}_k^\dagger+\mathrm{H.c.}\} \Big]\\
    &=& \hbar\sum_{j,k}G_{j,k}\hat{a}_j^\dagger\hat{a}_k+\frac{\hbar}{2}\sum_{j,k}\left(F_{j,k}\hat{a}_j^\dagger\hat{a}_k^\dagger+\mathrm{H.c.}\right),
\label{eq:linearized}
\end{eqnarray}
where \(g_0\) is the nonlinear coupling strength. By grouping the two sets of terms labeled by \(\delta_{(j\pm k\mp m-n)}\), we define parametric process terms $(F)$, and terms corresponding to Bragg-scattering four-wave mixing (BS-FWM), self- and cross-phase modulation, mode detunings, and dispersion $(G)$ \cite{gouzien_morphing_2020}, with H.c. denoting the Hermitian conjugate. 

We now consider the Heisenberg equations of motion in the quadrature basis for each mode: \(\mathbf{\hat{R}}(t)=\left( \mathbf{\hat{X}}(t)|\mathbf{\hat{P}}(t) \right)^{T}\), where \(\mathbf{\hat{X}}=(\hat{x}_{-N},...,\hat{x}_0,...,\hat{x}_N)^{T}\) and \(\mathbf{\hat{P}}=(\hat{p}_{-N},...,\hat{p}_0,...,\hat{p}_N)^{T}\), with \(\hat{x}_n = \frac{1}{\sqrt{2}}\left(a^{\dagger}_{n} + a_n\right)\) and \(\hat{p}_n = \frac{i}{\sqrt{2}}\left(a^{\dagger}_{n} - a_n\right) \). 
Using the input-output relations \cite{Gardiner_quantum_2010}, the input and output quadratures can be written in the Fourier domain as \(\mathbf{\hat{R}}_{\text{out}}(\omega)= S(\omega)\mathbf{\hat{R}}_{\text{in}}(\omega)\), where $S(\omega)$ is a conjugate-symplectic transfer function defined by: 
\begin{equation}
    S(\omega)=\sqrt{2\Gamma}\left( i\omega I + \Gamma - M\right)^{-1}\sqrt{2\Gamma} - I.
\label{eq:transfer_s}
\end{equation}
The mode interaction matrix \(M\in \mathbb{R}^{2N\times 2N}\) can be obtained directly from Eq.~(\ref{eq:linearized}): 
\begin{equation}
    M =
        \begin{pmatrix}
        \operatorname{Im}[G + F] & \operatorname{Re}[G - F] \\
        -\operatorname{Re}[G + F] & -\operatorname{Im}[G + F]^{T}
        \end{pmatrix}
        ,
\label{eq:m_matrix}
\end{equation}
with \(\Gamma\) as a diagonal matrix that contains the waveguide-cavity coupling rates $\Gamma_c$ and intrinsic losses $\Gamma_i$.  However, only the waveguide-coupled noise quadratures are physically accessible in experiments, leading to a reduction in squeezing and concomitant increase in anti-squeezing. We model the trace-out of the intrinsic loss channel with a beam-splitter of ratio $\Gamma_c/\Gamma$ after diagonalizing the squeezed supermodes (see S.2.) \cite{gouzien_hidden_2023, cui_high-purity_2021}.

Physically, one can view the Fourier analysis frequency \(\omega\) as the \textit{RF noise sideband frequency}, bridging the discrete frequency grid by creating noise suppression outside the resonant modes. The transfer function matrix $S(\omega)$ can therefore be diagonalized using the analytical Bloch-Messiah decomposition, \(S(\omega)=\mathcal{U}(\omega)\mathcal{D}(\omega)\mathcal{V}^{\dagger}(\omega)\). The diagonal matrix \(\mathcal{D}(\omega)\) contains the degree of anti-squeezing and squeezing of the corresponding supermodes in \(\mathcal{U}^\dagger(\omega)\) (Fig.~\ref{fig:schematic}). The columns of \(\mathcal{U}^\dagger(\omega)\) contain the linear coefficients of the frequency mode quadratures participating in the squeezed supermode, as they morph through $\omega$ \cite{gouzien_morphing_2020}. The generally complex nature of $\mathcal{U}(\omega)$ for both quadratures implies differences between the left and right sidebands. However, this asymmetry can be balanced out by implementing appropriate frequency unitaries \cite{Dioum_universal_2024, karnieli_variational_2025}, and we focus only on the supermode quadrature amplitudes in this article. We also note that in this work, \(S(\omega)\) is decomposed independently at each value of $\omega$, thus obtaining supermodes locally ordered by the $\omega$-dependent squeezing level, without the smooth morphing behavior.

We substitute steady-state spectra from the PDNLSE phase diagram into Eqs. (\ref{eq:linearized})-(\ref{eq:m_matrix}) and calculate the multimode squeezing and supermodes with the highest numerical squeezing, shown in Fig.~\ref{fig:phase_diagram}(a). All simulations are performed with the inclusion of intrinsic loss, assuming 99\% overcoupling: $\Gamma_i = 0.01\, \Gamma_c$, (see Supp. Mat.). We now contrast the squeezing spectra for two different states: below-threshold and stable soliton. 

\textit{Single- and two-mode squeezing}—Below threshold, we observe overlapping squeezing curves that cross at different analysis frequencies, with shifted peaks and dips for each squeezing/anti-squeezing pair, and an overall decay at larger $\omega$ due to the Lorentzian nature of the diagonal terms in \(S(\omega)\) (Fig.~\ref{fig:phase_diagram}(c)). Each curve overlaps in pairs with identical supermode amplitudes, and different phases (see S.3.), except for the degenerate supermode at $\mu=0$, corresponding to the highest squeezed supermode at $\Delta_\text{eff}=0$ (Fig.~\ref{fig:phase_diagram}(c) inset). This is emblematic of OPO phase-matching, where the single-mode and several two-mode squeezing processes compete via modal dispersion and nonlinear phase shifts from the pumps. In this simple scenario, the highest squeezed supermode for a given $\Delta_{\text{eff}}$ is decided entirely by the dispersion, as $\Delta_{\text{eff}}$ already accounts for nonlinear shifts, and so pairwise modes can similarly be maximized by tuning $\Delta_{\text{eff}}$ analytically (see S.4.).
This strong influence of the dispersion therefore indicates that lower dispersion can lead to high squeezing even at relatively higher $\omega$ values, provided that the nonlinear phase shifts can balance this. In the above threshold regime, these near-perfect phase-matched parametric processes at the zero-crossing modes naturally manifest as ``quantum" dispersive waves. In fact, qualitatively different phenomena can be seen above threshold, when nonlinear shifts, BS-FWM and even more parametric processes greatly influence the observed squeezing patterns.

\begin{figure}[]
\includegraphics[width=\linewidth]{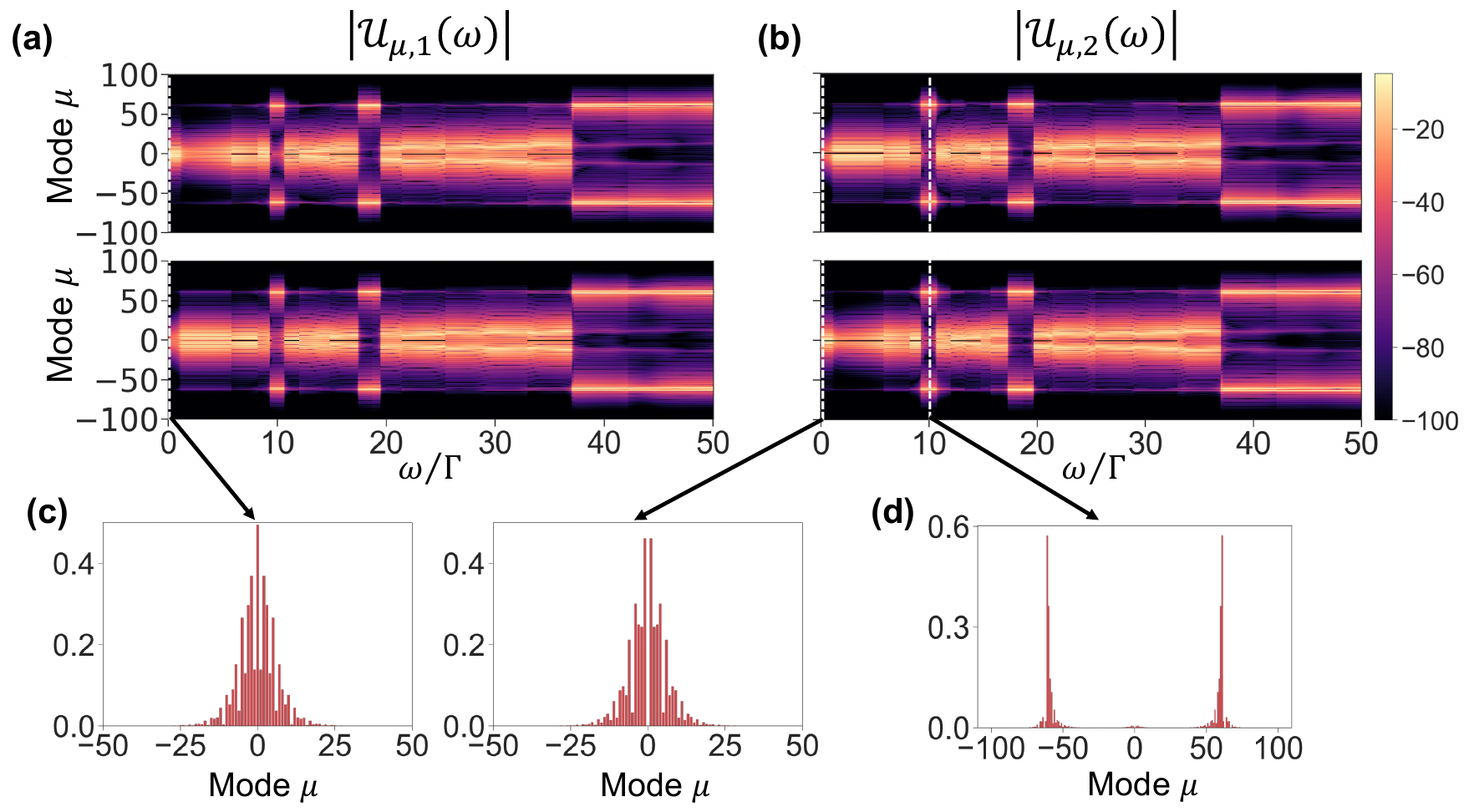}
\caption{\label{fig:single_two_mode} Degenerate mode alternation and pseudo-two-mode squeezing from quantum dispersive waves (QDWs). (a)-(b) Quadrature amplitudes (\(\mathbf{\hat{X}}\) above, \(\mathbf{\hat{P}}\) below) of the two highest squeezed supermodes, \(\mathcal{U}_1(\omega)\) and \(\mathcal{U}_2(\omega)\) for the stable soliton state in Fig.~\ref{fig:phase_diagram}(b), as a function of \(\omega/\Gamma\). The color-map shows the presence of QDWs around \(\mu=\pm61\) near analysis frequencies corresponding to the genuine crossings in Fig.~\ref{fig:phase_diagram}(b). The two highest squeezed supermodes  \(\mathcal{U}_1(\omega)\) and \(\mathcal{U}_2(\omega)\)  show complementary (bright vs dark) supermode amplitudes for the degenerate mode $(\mu=0)$ across the range of $\omega/\Gamma$, with their respective linecuts at $\omega=0$ shown in (c). (d) \(\mathcal{U}_2(10.0)\) supermode, whose amplitude is concentrated at the QDW peaks, closely resembling two-mode squeezing.} 
\label{fig:single_two_mode}
\end{figure}

\textit{Quantum dispersive waves}—Similar to below threshold, squeezing levels for stable solitons occur as nearly-overlapping pairs, with avoided-mode crossings at certain analysis frequencies \cite{patera_avoided_2025} (Fig.~\ref{fig:phase_diagram}(b)). With the inclusion of pumps from the now-bright modes of the comb, the squeezing spectrum is no longer defined  purely by $D_\mathrm{int}$ and $\Delta_\mathrm{eff}$, and also ceases to be peaked at $\omega=0$. The higher-order dispersion also plays a role in setting effective ``boundaries" for spreading quantum correlations to the azimuthal modes of the ring, as the modes beyond the zero-crossing are largely normal in dispersion \( (\hat{d}_\mathrm{int}<0) \). Fig.~\ref{fig:single_two_mode} shows the analysis frequency dependence of the first and second-highest squeezed supermodes, showing alternation in their respective amplitudes of the degenerate mode at $\mu=0$ at each $\omega$ value. As mentioned above, we expect abrupt swaps in the local supermode ordering at each $\omega$, corresponding with the genuine crossings in the squeezing levels. We thus cross into regimes where the highest squeezed supermode is localized near the zero-crossing points, which we term as quantum dispersive waves (QDWs). This is in analogy with classical soliton Cherenkov radiation \cite{PhysRevA.41.426, PhysRevA.51.2602, brasch_photonic_2016}, or dispersive waves in microcombs \cite{yi_single-mode_2017, kippenberg_dissipative_2018}. In our case, however, the classical comb (Fig.~\ref{fig:phase_diagram}(b) top inset) does not show such strong peaks at $\mu = \pm 60,61$, as is emphasized in the classical spectrum in log scale. In contrast, the quantum dispersive waves peak even in the linear scale, at large analysis frequencies. The continuous parameter $\omega$ therefore allows us to bridge the discrete frequency gaps between azimuthal modes and create phase-matching far away from resonance, where squeezing can now occur almost entirely concentrated at two low-dispersion modes. 

\begin{figure}[h!]
\includegraphics[width=\linewidth]{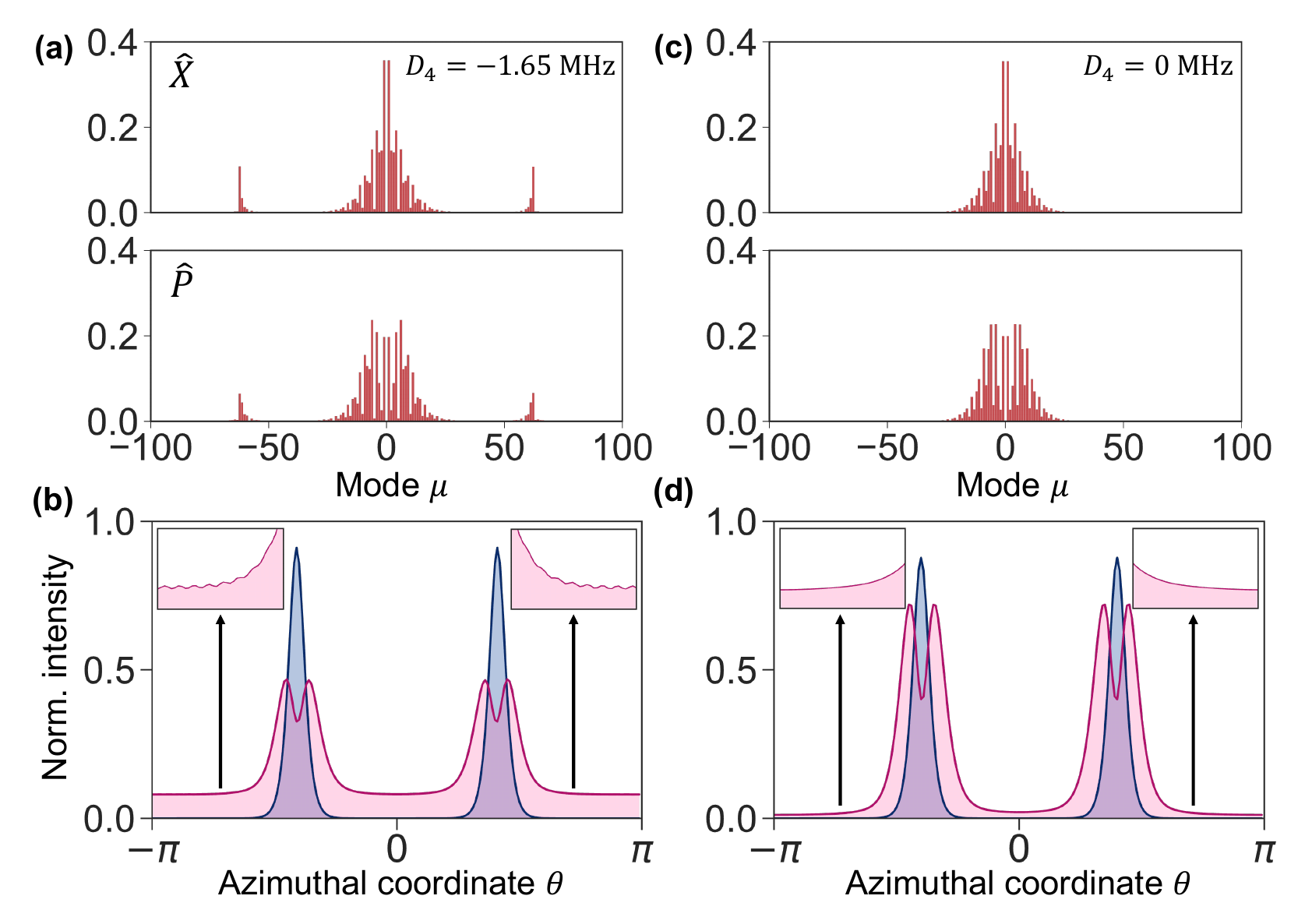}
\caption{\label{fig:epsart} Emergence (disappearance) of quantum dispersive waves (QDWs) in a quartic (purely quadratic) dispersion for the Stable Solitons state in Fig.~\ref{fig:phase_diagram}(b). (a) Quadrature amplitudes of the highest squeezed supermode showing QDWs around the zero crossings of a quartic integrated dispersion. (b) Temporal envelope of the squeezed state \(\langle\hat{a}^{\dagger}(\theta)\hat{a}(\theta)\rangle\) (pink) plotted with the normalized mean field intensity (blue) corresponding to (a). Insets: The temporal envelope showcases an elevated noise background with fast oscillations along the azimuthal coordinate. (c)-(d) same as (a)-(b) but with a quadratic dispersion. The QDWs and temporal background oscillations do not appear.}
\label{fig:qdws}
\end{figure}

For further evidence of the occurrence of quantum dispersive waves, we construct the temporal noise envelopes of these comb states following the procedure of \textcite{guidry_multimode_2023}. We construct the photon number operator at each azimuthal coordinate by inverting Fourier transforms over the $\omega$ and $\mu$ axes (see S.7.). This provides a real-valued temporal envelope of the intra-cavity noise. We even obtain a double-peaked shape in Fig.~\ref{fig:qdws}, agreeing with predictions for soliton crystals and synchronously pumped OPOs \cite{guidry_multimode_2023, de_valcarcel_multimode_2006, patera_quantum_2009}. To observe the effect of the QDWs on the temporal noise envelope, we remove the effects of quartic dispersion, and the zero crossings thereof, and simulate the PDCS classically in conditions which deviate from the originally proposed constraints. Although going against the pump placement requirement, it provides direct insights into how the supermodes and noise envelope shift with the inclusion of QDWs (Fig.~\ref{fig:qdws}). We see an increased noise floor with fast oscillations away from the azimuthal positions of the pulses, which is completely absent when $D_4$ is set to zero, further solidifying the occurrence of quantum dispersive waves even in the absence of classical dispersive wave signatures in the mean field. 

\textit{Discussion}—In summary, we theoretically present the highly multimode squeezing of pure-Kerr parametrically driven CSs in dual-pumped microresonators. By employing experimentally feasible parameters and the analytical Bloch-Messiah decomposition, we report high squeezing across the different dynamical regions of the PDNLSE phase diagram, bridging the classical and quantum dynamics of these solitons. We further identify quantum dispersive waves above threshold, and verify them with the temporal noise envelope. Our work not only provides one of the first quantum analyses of these newly-discovered cavity solitons, but also promotes their potential for generating large-scale quantum correlations. Particularly, the occurrence of pseudo-two mode squeezing even in the presence of complex classical comb states implores further investigation for several applications in quantum metrology, quantum information processing and sensing.

\textit{Acknowledgments}—We acknowledge M. Erkintalo, S. Shamailov, and G. Moille for helpful discussions about the parametrically-driven nonlinear Schr\"odinger equation and its classical dynamics. R.R.M. gratefully acknowledges support from the NSF MathQuantum RTG (NSF award \#DMS-2231533) and Daniel Serrano. R.R.M. and S.K.S. thank Samarth Sriram and Antoine Henry for stimulating discussions. P.S. was supported by NIST (\#60NANB24D106) and the UMBC Quantum Science Institute. This work was supported by grants from the Naval Air Warfare Center, Aircraft Division and from the National Science Foundation (\# 2326792). A.D. was supported by an NSF CAREER grant (\# 2340835).



\bibliography{apssamp, avik_zotero_library}

\clearpage

\setcounter{secnumdepth}{3}
\setcounter{section}{0}
\renewcommand{\thesection}{S.\arabic{section}}

\setcounter{equation}{0}
\renewcommand{\theequation}{S\arabic{equation}}

\setcounter{figure}{0}
\renewcommand{\thefigure}{S\arabic{figure}}

\setcounter{table}{0}
\renewcommand{\thetable}{S\arabic{table}}

\begin{center}
    \textbf{\large Supplementary Material}
\end{center}

\section{Parametrically-driven Nonlinear Schr\"odinger Equation}

The dynamics of parametrically driven cavity solitons can be simulated using the extended Lugiato-Lefever equation (LLE)  \cite{moille_parametrically_2024, Taheri2017}: 

\begin{equation}
\begin{aligned}
t_R\frac{\partial E(t,\tau)}{\partial t}
=
\left[
-\Gamma + i\left(\gamma L |E|^2 - \delta_0\right)
- i \sum_{k>1}\frac{D_k}{k!}(\mu)^k
\right] E\\
+ \sqrt{\theta_+}E_+ e^{-i\Omega_p \tau + i b_+ t}
+ \sqrt{\theta_-}E_- e^{i\Omega_p \tau - i b_- t},
\end{aligned}
\label{eq:LLE}
\end{equation}

\noindent where $t_R$ is the roundtrip time, $\delta_0$ is the central detuning, \(\Omega_p\) is the (angular) frequency separation between \(\omega_\pm\) and the central PDCS frequency \(\omega_0\), $b_\pm$ are coefficients depending on detunings of \(E_\pm\) and dispersion values, and \(\theta_\pm\) are the coupling coefficients of each driving field. The integrated dispersion $D_{\text{int}}$ is defined such that it accounts for the detuning of a cavity frequency mode from a regularly spaced frequency grid: 
\begin{equation}
\begin{aligned}
\omega_{\mu} = \omega_0 + \mu D_1 + D_{\text{int}}(\mu) = \omega_0 + \mu D_1 + \sum_{k>1}\frac{D_k}{k!}(\mu)^k.
\end{aligned}
\label{eq:D_int}
\end{equation}

The extended LLE (Eq.~\ref{eq:LLE}) describes the dynamics of a bichromatically driven resonator with pumps placed near the zero-dispersion crossings to ensure the linear phase matching of the degenerate four-wave mixing (FWM) process. As described in Ref. \cite{moille_parametrically_2024}, one can engineer the dispersion such that the intracavity fields \(E_\pm\) at the pump frequencies are homogeneous and stationary. In this scenario, the slowly varying envelope can be separated into the following ansatz: 

\begin{equation}
\begin{aligned}
E(t,\tau) &= E_0(t,\tau) \\
&\quad + E_{+} e^{-i\Omega_p\tau+i b_+ t}
+ E_{-} e^{i\Omega_p\tau-i b_- t},
\end{aligned}
\label{eq:anszat}
\end{equation}
Furthermore, assuming that the PDCS spectrum \(E_0\) does not overlap with the pump fields, one can substitute Eq.~\ref{eq:anszat} into the extended LLE (Eq.~\ref{eq:LLE}) to isolate the PDCS evolution $E_0(t,\tau)$ and obtain (after normalizing) the parametrically driven nonlinear Schr\"odinger equation (PDNLSE) shown in Eq. (1) in the main text \cite{moille_parametrically_2024}. While PDNLSE does not consider the dynamics of the pumps and instead only their parametric processes induced by the parametric drive \(\nu = 2iE_+ E_-\) and nonlinear shift included in the $\Delta_{\text{eff}}$, we do include the interactions of the stationary pumps when constructing the mode interaction matrix \(M\) for the multimode squeezing analysis (defined in Eqs. (2)-(4)). 

The parameters used to simulate the PDCS spectrum were taken from a recent experimental realization detailed in Ref. \cite{moille_parametrically_2024}. A $23\,\mu$m radius, $1$ THz free spectral range (FSR) $\text{Si}_3\text{N}_4$ resonator is considered with finesse $\mathcal{F}=\pi/\Gamma=3000$, where $\Gamma$ is the total loss per roundtrip. A quartic dispersion profile is assumed with $D_2/2\pi = 0.5~$GHz, $D_4/2\pi = -1.645~$MHz. We simulate the PDNLSE for 200 modes: $\mu=-100$ to $\mu=+99$. These dispersion parameters are not unique, in the sense that different combinations of $D_2, D_4, \Delta_{\text{eff}} \text{ and }\nu $ can also lead to sustained stable solitons. The nonlinear coefficient $\gamma$ is taken to be the typical value for $\text{Si}_3\text{N}_4$,  $1\, \rm W^{-1}m^{-1}$.

The transformations used to convert the extended LLE (Eq.~\ref{eq:LLE}) to the normalized PDNLSE are:
$\tau \rightarrow \tau \sqrt{2\Gamma/|\beta_2|L}$, $t \rightarrow \Gamma t /t_R$, $\Omega_p \rightarrow \Omega_p \sqrt{|\beta_2|L/2\Gamma}$, $E \rightarrow E\sqrt{\gamma L/\Gamma}$ \cite{coen_universal_2013, moille_parametrically_2024}. The normalized integrated dispersion is given in the main text following Eq. (1).

\section{Addition of intrinsic loss}
The addition of intrinsic loss in the squeezing calculations can be achieved by following Ref. \cite{cui_high-purity_2021} and Appendix E of Ref. \cite{gouzien_hidden_2023}. Here, one can treat intrinsic loss as a virtual (input) waveguide with loss $\Gamma_i$, in this case intrinsic loss, mixing with the input quantum modes through a beam-splitter with a reflectivity of \(\Gamma_c/\Gamma\) before decomposing into squeezing values \(\mathcal{D}^2(\omega)\) and their respective supermodes \(\mathcal{U}^\dagger(\omega)\), where \(\Gamma_c\) is the coupling loss. After the decomposition, the second beam-splitter mixes the output quantum modes with the virtual waveguide once again to obtain the squeezing spectrum with intrinsic loss:

\begin{equation}
    \mathcal{D}^2 _{\text{intrinsic}} (\omega) = \frac{\Gamma_i}{\Gamma} + \frac{\Gamma_c}{\Gamma}\mathcal{D}^2(\omega).
\label{eq:intrinsic}
\end{equation}

We note that \(\mathcal{D}(\omega)\) in the main paper corresponds to \(\mathcal{D}_{\text{intrinsic}}(\omega)\) already accounting for intrinsic losses, and we only exchanged their notation. Moreover, we highlight that the supermodes remain unchanged under intrinsic loss and only the squeezing spectrum is altered. Eq.~\ref{eq:intrinsic} can be verified in the limit of \(\Gamma_i \rightarrow 0\), where the beam splitter reduces to the decomposition shown in Fig. (1) in the main text. 

\begin{figure*}[]
\includegraphics[width=\linewidth]{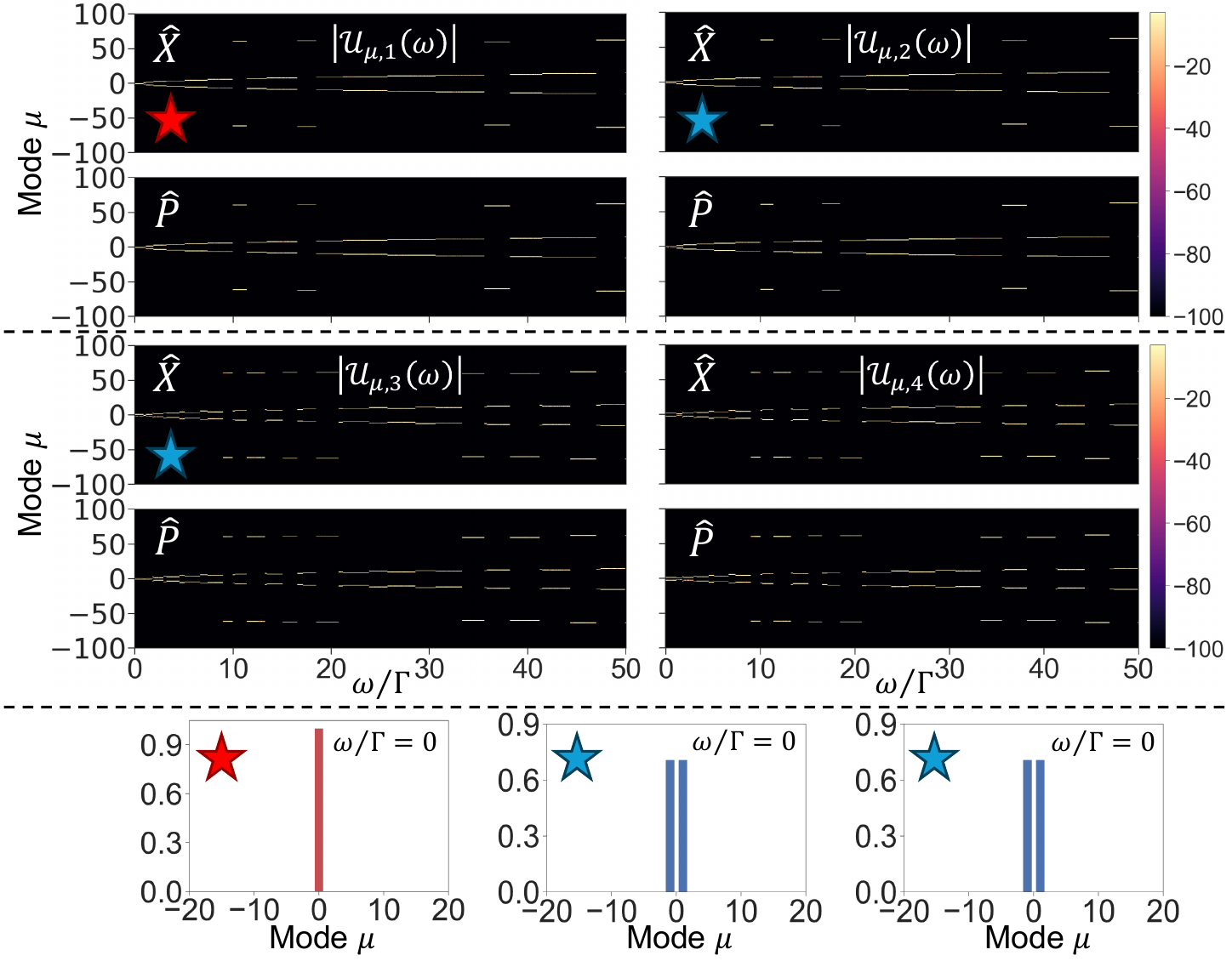}
\caption{\label{fig:epsart} Higher-order supermodes for a below-threshold state for parameters: \((\Delta_{\text{eff}}=0,\nu=0.95)\). Top and middle: \(\hat{\mathbf{X}}\) and \(\hat{\mathbf{P}}\) quadratures of \(\mathcal{U}_1(\omega)\) -- \(\mathcal{U}_4(\omega)\) as a function of \(\omega/\Gamma\). Each heatmap pair (\(\mathcal{U}_1(\omega)\), \(\mathcal{U}_2(\omega)\) and \((\mathcal{U}_3(\omega)\), \(\mathcal{U}_4(\omega)\)) manifests degeneracy for all \(\omega/\Gamma\) values except at \(\mathcal{U}_1(0)\). 
Bottom: quadrature amplitudes of supermodes \(\mathcal{U}_1(0)\) -- \(\mathcal{U}_3(0)\). While the pairwise degeneracy for all analysis frequencies is discontinued due to the single-mode squeezing emergence, there is still degeneracy in the supermodes at \(\omega/\Gamma=0\), which shifted to the next higher-order supermodes. In the squeezing spectra, the squeezing levels overlap for every pairwise degeneracy, while showing a disruption for the highest squeezing level (single-mode squeezing).     
}
\label{fig:bt_supermodes}
\end{figure*}

\section{Higher-order supermodes for below-threshold states}

In Fig. 2 from the main text, we obtained the squeezing spectrum for a below-threshold (BT) state with phase-space parameters \((\Delta_\text{eff}=0.0,\nu=0.95)\), where we also showed the highest squeezed supermodes correspond to single-mode squeezing. Here, we show that the supermodes come in pairs, leading to squeezing overlap except at the analysis frequency corresponding to the single-mode supermode. Fig.~\ref{fig:bt_supermodes} shows the \(\hat{\mathbf{X}}\) and \(\hat{\mathbf{P}}\) higher-order supermodes as a function of \(\omega/\Gamma\) for the below-threshold state in the main text as well as the quadrature amplitudes of supermodes \(\mathcal{U}_1(0)\)-\(\mathcal{U}_3(0)\). Every pair of supermodes in the heatmaps exhibits degeneracy for every \(\omega/\Gamma\) except at \(\mathcal{U}_1(0)\), which corresponds to the single-mode squeezing (Fig.~\ref{fig:bt_supermodes}, red star). While the pairwise degeneracy of \(\mathcal{U}(\omega)\) is discontinued at \(\omega/\Gamma=0\), the degeneracy at such analysis frequency is shifted to the next highest supermode (see Fig.~\ref{fig:bt_supermodes} red and blue stars to observe the degeneracy at \(\omega/\Gamma=0\)), where the supermodes \(\mathcal{U}_2(0)\) and \(\mathcal{U}_3(0)\) continue the pairwise degeneracy. This discontinuity in the degeneracy of \(\mathcal{U}(\omega)\), together with the corresponding degeneracy shift of the supermodes at a particular \(\omega/\Gamma\), indicates the emergence of single-mode squeezing across different phase-space parameters within the below-threshold region. Lastly, the supermodes go from single- to two-mode squeezing as the analysis frequency varies, with some two-mode squeezing manifested at the dispersion zero-crossings corresponding to analysis frequencies with genuine crossings in the squeezing spectrum (see Fig. 2 in main text).  

\section{Squeezed supermodes using perfect mode-phase-matching}

The Below-Threshold states also allow us to investigate the interplay between dispersion and effective detuning, leading to perfect mode-phase-matched squeezing. First, under below-threshold pumping, no coherent steady-state field is sustained, and hence the classical steady-state amplitudes are taken to be zero, simplifying the interaction Hamiltonian \(H_{\text{int}}\) for quantum analysis. Following the definitions of \(F\) and \(G\) in the main text and Ref.~\cite{gouzien_morphing_2020, gouzien_hidden_2023}, we can analytically construct the mode interaction matrix \(M\): The non-degenerate FWM interactions induced by the parametric drive \(\nu\) can be directly added to the corresponding matrix-elements of \(F\), while the diagonal of \(G\) only contains the linear and nonlinear (XPM and SPM) shifts ensued from the dispersion and effective detuning \(\Delta_{\text{eff}}\). 

\begin{figure}[h!]
\includegraphics[width=\linewidth]{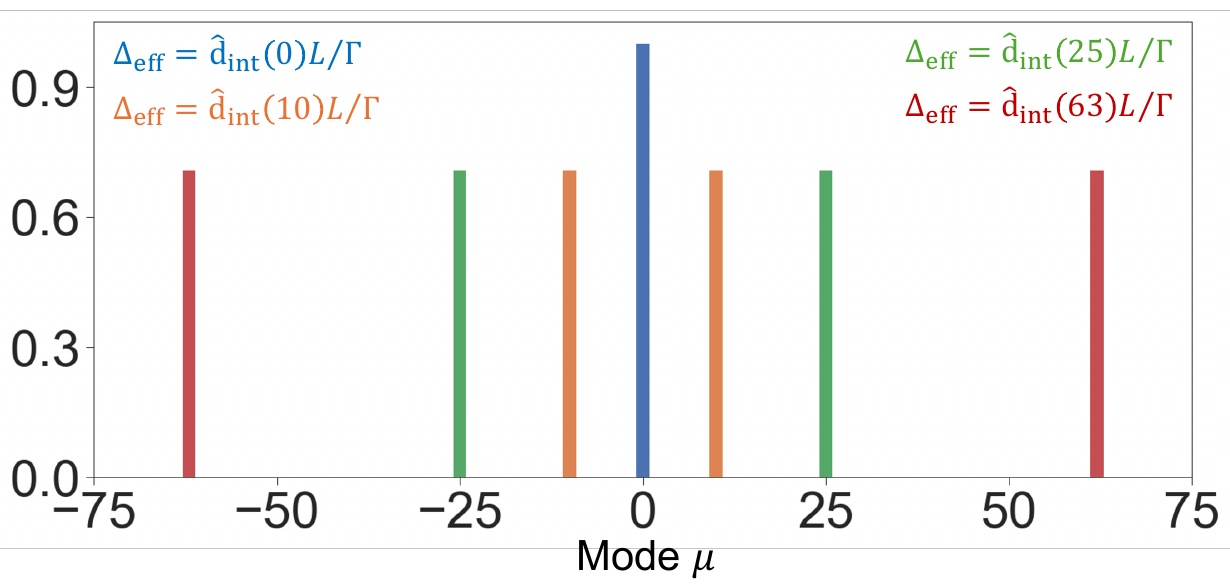}
\caption{\label{fig:epsart} Controllable single- and two-mode squeezing in BT states using perfect mode-phase-matching condition at \(\omega/\Gamma=0\). Given the simplified interactions in the mode interaction matrix \(M\) at below-threshold conditions, it is possible to maximize the squeezing at a specific quantum mode, either single- or two-mode, following phase-matching condition with \(\Delta_\text{eff}\) and \(\hat{d}_\text{int}(\mu)\) by using Eq. ~\ref{eq:deltaeff} across different modes \(\mu\). This controllable squeezing, in particular near the zero-dispersion crossings, suggests the combined roles of effective detuning, dispersion, and analysis frequencies in the squeezing and supermodes in the Below-Threshold and above-threshold states.}
\label{fig:controlsq}
\end{figure}

As detailed in the main text and shown in Fig.~\ref{fig:bt_supermodes}, the supermodes in the Below-Threshold states manifest pairwise degeneracy, leading to two-mode and single-mode squeezing. In particular, the two-mode squeezed supermodes are dependent on the quantum modes as well as the dispersion, as for some analysis frequencies there are abrupt jumps to the zero-dispersion crossings (\(\mu=\pm63\). Given that the parametric drive is fixed under OPO-threshold, the effective detuning \(\Delta_{\text{eff}}\) and the dispersion \(\hat{d}_{\text{int}}(\mu)\) (defined in the main text), whose values vary as a function of the frequency mode, act as ``knobs" to control the modes with the highest squeezing. Here, \(\hat{d}_{\text{int}}(\mu)\) is already determined by the cavity's condition for PDCS formation, so we devise an analytical expression to maximize the squeezing at a particular quantum mode by ensuring perfect phase-matching condition between \(\Delta_{\text{eff}}\) and \(\hat{d}_{\text{int}}(\mu)\):

\begin{equation}
    \Delta_\text{eff} = \frac{\hat{d}_\text{int}(\mu)L}{\Gamma}.
\label{eq:deltaeff}
\end{equation}

To demonstrate perfect mode-phase-matched squeezing, we use Eq.~\ref{eq:deltaeff} into the Hamiltonian construction and obtain the highest squeezed supermodes. In Fig.~\ref{fig:controlsq}, we plot the supermodes with the largest squeezing at \(\omega/\Gamma=0\) for detuning values chosen to enforce phase matching between the interacting mode and the dispersion profile. We note that the \(\Delta_\text{eff}\) are negative due to the anomalous dispersion regime. Using the mode-phase matching condition in Eq.~\ref{eq:deltaeff}, we show that it is possible to maximize a particular supermode profile, either single- or two-mode squeezing, and to control the quantum mode at which such squeezing occurs. In addition to highlighting the fine-tuning of \(\Delta_\text{eff}\) for controllable single- and two-mode squeezing, this insight allows us to investigate the interplay between \(\hat{d}_\text{int}(\mu)\) and \(\Delta_\text{eff}\) in the below-threshold supermodes, and draw intuition for the above threshold PDCSs. For instance, the two-mode squeezing supermodes near the \(\hat{d}_\text{int}(\mu)\) zero-crossings even when pumped below threshold already hints at the emergence of quantum dispersive waves shown in above threshold states due to \(\Delta_\text{eff}\), \(\hat{d}_\text{int}(\pm\mu)=0\), and the analysis frequencies \(\omega/\Gamma\).     

\section{Multimode squeezing for other above-threshold PDCS}
\begin{figure}[h!]
\includegraphics[width=\linewidth]{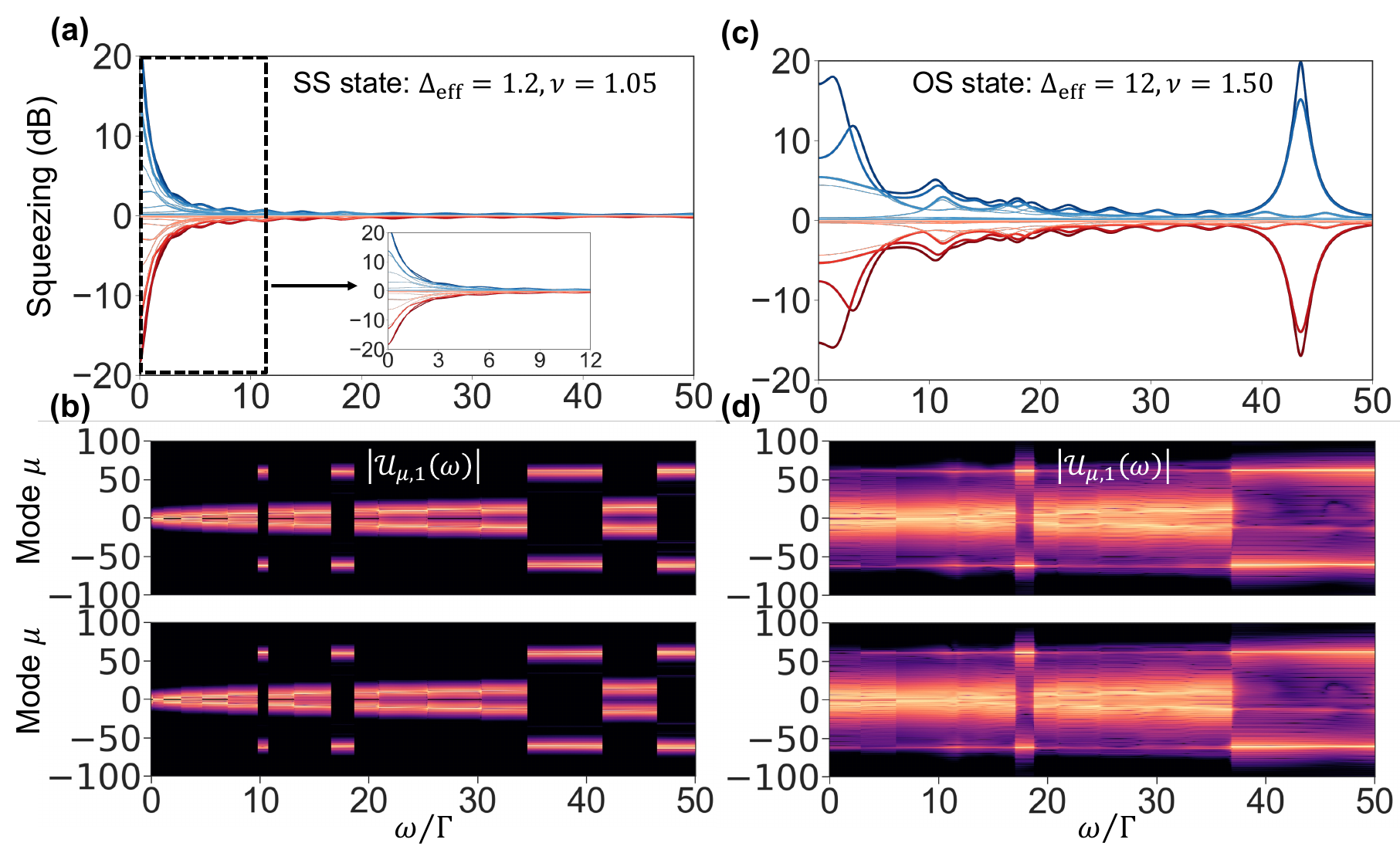}
\caption{\label{fig:epsart} Multimode squeezing analysis for other soliton states in the PDNLSE bifurcation diagram. (a)-(b) shows the squeezing spectrum and quadrature coefficients of \(\mathcal{U}_1(\omega)\) for a stable soliton (SS) state with phase-space parameters \((\Delta_{\text{eff}}=1.2, \nu = 1.05)\). (c)-(d) same as (a)-(b) but for an oscillatory solitons (OS) state with \((\Delta_{\text{eff}}=12, \nu = 1.50)\).}
\label{fig:other_states}
\end{figure}

Here we show additional multimode squeezing analysis for different soliton states in the above-threshold region. Fig. 2 in the main text presents the classical bifurcation diagram of the PDNLSE, yielding different soliton states including stable solitons, oscillatory solitons (OS), and Turing patterns (TP). Although the paper focuses mainly on multimode squeezing for a stable soliton state with phase-space parameters \((\Delta_{\text{eff}}=12, \nu = 1.05)\), we calculated the highest squeezing across different states. Fig.~\ref{fig:other_states} shows the squeezing spectrum and the supermodes coefficients \(\hat{\mathbf{X}}\) and \(\hat{\mathbf{P}}\) of \(\mathcal{U}_1(\omega)\) for another stable solitons \((\Delta_{\text{eff}}=1.2, \nu = 1.05)\) and oscillatory solitons \((\Delta_{\text{eff}}=12, \nu = 1.05)\). The stable solitons, pumped just slightly above threshold with a small \(\Delta_\text{eff}\) and close to the below-threshold state in Fig. 2, displays less multimode behavior compared to that of Fig. 2 and instead its supermodes (Fig.~\ref{fig:other_states}) reassemble an ``amplified" version of the below-threshold state supermodes (Fig.~\ref{fig:bt_supermodes}, top and middle). Conversely, the oscillatory solitons' squeezing spectrum displays similar features to that of Fig. 2 with more asymmetrical excited supermodes owing to the oscillatory nature of the frequency comb.   

\section{Supermodes for above-threshold PDCS with quadratic dispersion}

\begin{figure}[]
\includegraphics[width=\linewidth]{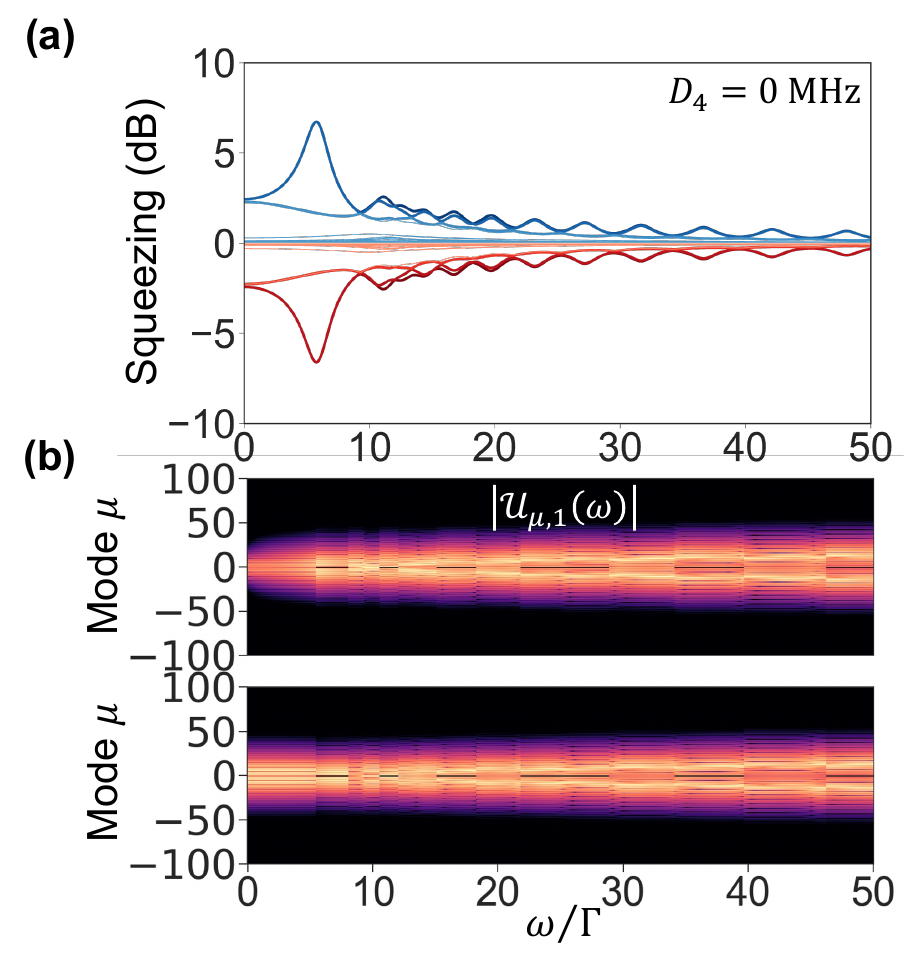}
\caption{\label{fig:epsart} Squeezing spectra and quadrature coefficients of \(\mathcal{U}_1(\omega)\) for a PDCS in a purely quadratic dispersion. When compared to the quartic dispersion case in Fig. 2 and Fig. 3 in the main text, the purely quadratic case lacks the complex interactions of \(\Delta_\text{eff}\) and zero-dispersion crossings as well as the manifestation of the quantum dispersive waves.}
\label{fig:noD4}
\end{figure}

In Fig. 4 of the main text, we simulated the formation of a PDCS in a purely quadratic (\(D_4 = 0\)) dispersion to demonstrate the role of zero-dispersion crossings in the emergence of quantum dispersive waves. Fig.~\ref{fig:noD4} shows the squeezing spectra as well as the \(\hat{\mathbf{X}}\) and \(\hat{\mathbf{P}}\) supermode heatmaps for \(\mathcal{U}_1(\omega)\). Clearly, the squeezing spectra lacks the highly-entangled levels at large \(\omega/\Gamma\) values coming from the interplay between the analysis frequencies and the quartic dispersion in Fig. 2, which is also manifested in the supermodes by the disappearance of the quantum dispersive waves at \(\mu=\pm63\). 

\section{Temporal noise shape calculation}

A key feature of dispersive waves in dissipative Kerr solitons (DKSs) is the emission of soliton Cherenkov radiation \cite{brasch_photonic_2016, kippenberg_dissipative_2018}. The presence of higher-order dispersion leads to an interference between the soliton spectrum placed at the center and the sub-comb formation near the zero-dispersion crossings, hence producing rapid oscillatory waves in the temporal domain. We showed in the main text (see Fig. 4) the emergence of an analogous behavior in the PDCS squeezed supermodes—quantum dispersive waves (QDWs). To further confirm this phenomenon, we calculated the temporal noise shape of the cavity fluctuations following Ref. \cite{guidry_multimode_2023}. This noise shape corresponds to the photon number operator for each azimuthal mode in the microresonator. 

We start by defining column vectors of the intracavity fluctuations and the input fluctuations: 

\begin{eqnarray}
\bar{a}(\omega) &=
\begin{pmatrix}
\hat{a}_1(\omega) \\
\vdots \\
\hat{a}_N(\omega) \\
\hat{a}_1^\dagger(-\omega) \\
\vdots \\
\hat{a}_N^\dagger(-\omega)
\end{pmatrix}, \\[10pt]
\bar{b}_\text{in}(\omega) &=
\begin{pmatrix}
\hat{b}_{\text{in},1}(\omega) \\
\vdots \\
\hat{b}_{\text{in},N}(\omega) \\
\hat{b}_{\text{in},1}^\dagger(-\omega) \\
\vdots \\
\hat{b}_{\text{in},N}^\dagger(-\omega)
\end{pmatrix}.
\end{eqnarray}
The intracavity and input fluctuations can be related by a transfer function \(\bar{a}(\omega)=Q(\omega)\bar{b}_\text{in}\) in the creation/annihilation operator basis: 

\begin{align}
Q(\omega) &= \left[i\omega + \Gamma- \hat{O}^\dagger\textit{M}\hat{O}\right]^{-1}\sqrt{2\Gamma}, \\
          &= \begin{pmatrix}
                Q_1 & Q_2 \\
                Q_3 & Q_4
             \end{pmatrix}_{2N \times 2N}, \\
\hat{O} &= \frac{1}{\sqrt{2}} \begin{pmatrix}
                \mathbb{I} & \mathbb{I} \\
                -i\mathbb{I} & i\mathbb{I}
             \end{pmatrix}_{2N \times 2N}. 
\end{align}
where $\hat{O}$ is the necessary basis transformation from quadratures to creation/annihilation operators. After defining the matrix $\Theta$ with elements \(\Theta_{lm}=\sum_{l,m}^N e^{-i(l-m)\theta}\), we take the expectation value of the intracavity modes (in time domain) to arrive at the following expression for the azimuthal photon number operator: 
\begin{equation}
    \left \langle\hat{a}^\dagger(\theta)\hat{a}(\theta) \right \rangle = \int_{-\infty}^{\infty} d\omega \left \langle \Theta, Q_2 \cdot Q_2^\dagger \right \rangle (\omega).
\label{eq:photon}
\end{equation}
where $\langle A,B\rangle =\sum {A}^*_{ij} B_{ij}$ is the Frobenius inner product. 



\end{document}